\documentclass[a4paper]{IEEEtran}

\usepackage[utf8]{inputenc}
\usepackage[noadjust]{cite}
\usepackage{url}
\usepackage{color}
\usepackage{subcaption}
\usepackage{graphicx}
\usepackage{amstext}

\title{Evaluating socio-economic state of a country analyzing airtime credit and mobile phone datasets}

\author{
\begin{tabular}{ccccc}
  Thoralf Gutierrez & \quad & Gautier Krings & \quad & Vincent D. Blondel\\
  ICTEAM && Real Impact Analytics && ICTEAM\\
  Université catholique de Louvain && && Université catholique de Louvain\\
  thoralf.gutierrez@student.uclouvain.be && && vincent.blondel@uclouvain.be
\end{tabular}
}

\begin{document}

\maketitle

\begin{abstract}
Reliable statistical information is important to make political decisions on a sound basis and to help measure the impact of policies. Unfortunately, statistics offices in developing countries have scarce resources and statistical censuses are therefore conducted sporadically. Based on mobile phone communications and history of airtime credit purchases, we estimate the relative income of individuals, the diversity and inequality of income, and an indicator for socioeconomic segregation for fine-grained regions of an African country. Our study shows how to use mobile phone datasets as a starting point to understand the socio-economic state of a country, which can be especially useful in countries with few resources to conduct large surveys.
\end{abstract}

\begin{IEEEkeywords}
mobile phones, ICTs for development, socio-economic level, wealth distribution, poverty map, communities, socio-economic segregation.
\end{IEEEkeywords}

\section{Introduction}

Statistical information is important to foster awareness of the social, demographic, economic and environmental conditions of a country. They help forge opinions, make political and evidence-based decisions on sound foundations. They are also useful to help allocate resources. Unfortunately, many developing countries do not have much up-to-date statistical information about the state of their population. Surveys are onerous, they are not conducted very often. For example, the last population census in Angola was conducted in 1970 \cite{tatem}. Even when they are conducted, their reliability is questionable (see Appendix). Indeed, the figures of official reports usually involve a great deal of guesswork \cite{jerven}, as individuals and enterprises are less likely to be officially registered and to keep formal records of their economic activities. Up-to-date reports are of primary importance in Côte d'Ivoire, as the recent civil war has changed the face of the country.\\

With ubiquitous sensors like mobile phones, we have the opportunity to get fact-based figures at the individual level, and aggregate them to get local information across the country. Furthermore, the logs of mobile phone activity are generated in real-time and can therefore be used as an early alarm system, instead of waiting for the next survey. As these logs are already being generated and stored for legal purposes, they are virtually free compared to censuses, and could thus be of great use to developing countries that have few resources.\\

In this work, we explore mobile phone logs and airtime credit purchases to identify and extract proxies for wealth in Côte d'Ivoire.\\

Our study is based on 2012 logs from a large mobile operator in Côte d'Ivoire that have a client base representative of the population. The dataset contains the Call Detail Records for calls and text messages arising from clients. They have a timestamp, an ID for the initiator, an ID for the receiver and an identifier for the cell tower used. The dataset also contains the timestamp and amount of every airtime credit purchase made by every client.\\


Firstly, we summarize results available in the literature on similar questions. Secondly, we explore the airtime credit purchases of users and observe behavioral patterns in order to identify a proxy for wealth. We then map the average of the proxy with its variation and distribution inequality. Finally, we look at the data from a social networking angle. We detect the underlying communities in the network and observe a degree of wealth homophilia for the members of the same community.


\section{Related work}

As mobile phone datasets have been increasingly made available to research teams, links between mobile phone usage and socio-economic levels have been studied by many.\\


Eagle et al. have developed a new metric to capture the social diversity of communication ties within an individual's social network \cite{social_diversity}, where high diversity scores imply that an individual splits his time more evenly among social ties. The authors showed that, in the United Kingdom, social diversity scores were strongly correlated with the government's Index of Multiple Deprivation rank. They showed the same result for the number of structural holes \cite{burt} spanned by an indivudal's links.\\

Wang et al. proposed a measure of dyadic reciprocity that captures the degree of communicative imbalance between two nodes \cite{wang}. The index is minimal when two nodes have exactly the same probability of calling each other, and increases monotonically with the difference between probabilities. Researchers at UN Global Pulse explained how this reciprocity index had been linked to higher socio-economic levels \cite{kirkpatrick_strata}.\\

Blumenstock et al. \cite{blumenstock2010} inferred household expenditures of a small, but representative, set of users from a phone survey in Rwanda. They then analyzed the correlations with mobile phone usage and household expenditures of the queried users. They found correlations, mainly with the number of international calls, and the number of different districts contacted. They also found a positive correlation with the average airtime credit purchase, but they insist that they did not use a robust approach and hope to do so in future work. Blumenstock \& Eagle \cite{blumenstock2012} did a similar analysis, but compared the usage of the lower and upper quartiles of people, based on their estimate of socio-economic level. They showed that richer people had a higher maximum value of airtime credit purchases than poorer people. They did not give numbers for the average of airtime credit purchases.\\

Soto et al. \cite{soto} and Frias-Martinez et al. \cite{frias-martinez} have focused on mobility variables. They looked at many of them, and reported the following top features~: the number of cell tower used per week, the maximum distance traveled in a week and the weekly radius of gyration (i.e. the typical range of a user trajectory \cite{gonzalez}). Both were able to predict socio-economic classes of neighborhoods in their city of study.\\

Smith et al. \cite{smith} and Mao et al. \cite{mao} also worked on Côte d'Ivoire. Smith et al. found a strong correlation between census data on poverty and the following variables~: gravity residuals or the sum of differences between observed and expected flows using a gravity model, the number of calls and total duration, the social diversity as defined by Eagle~et~al., and introversion or the ratio of self-flow to total flow.\\

Mao et al. did a similar analysis, but based on different census data. They observed the highest correlation with socio-economic levels using the ratio of weighted out-degree to the sum of weighted out and in-degree.\\

Unfortunately, we argue in the Appendix that the census data from Côte d'Ivoire, on which the authors based their models, are questionable. A hint of this is given by the difference between their poverty maps. Furthermore, the censuses are at least four years older than the authors' datasets, and were conducted before the civil war.

\section{Purchases}

\begin{figure}
  \centering
  \includegraphics[width=\linewidth]{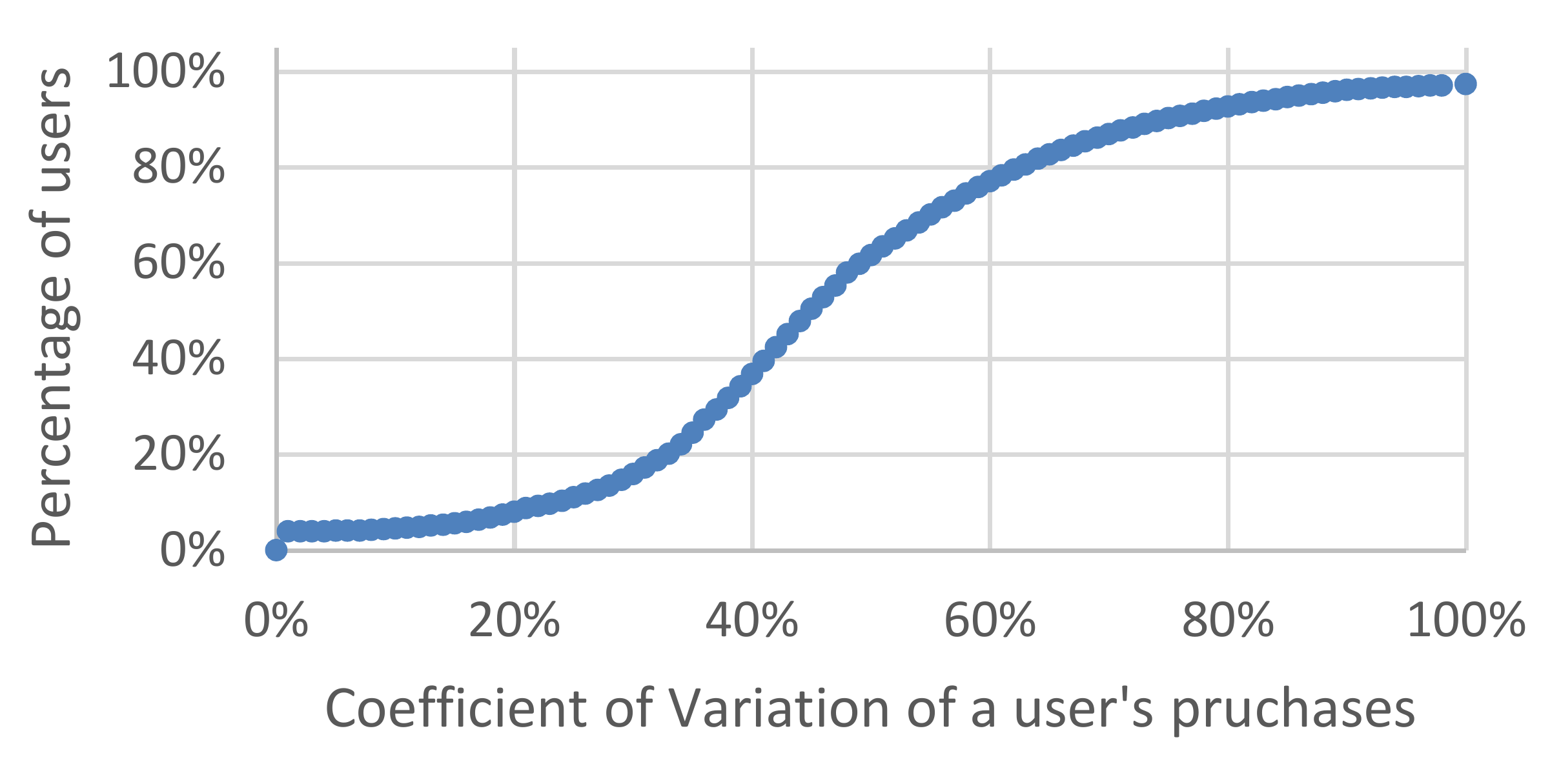}
  \caption{Cumulative frequency analysis of Coefficients of Variation of each user's purchases}
  \label{fig:stable_cv}
\end{figure}

We start by analyzing airtime credit purchases. We first look at each individual's behavior, by quantifying the variation in purchase amounts of each user with their Coefficient of Variation (CV). It is equal to the ratio of the standard deviation $\sigma$ to the average $\mu$. $$\text{CV} = \sigma/\mu$$ This variability measure is independent of the series average, and translates into the fact that it would be as easy or difficult for someone to make a purchase twice the size of what he is used to, whatever the amount he is used to. Figure \ref{fig:stable_cv} shows a cumulative frequency analysis of CVs. We see that 80\% of the users have a CV under 62\%, and 97\% of the users have one under 100\%. As a reference, the CV of the series of all purchases of every client is 238\%. The amounts purchased by most people appear to be stable. To simplify the data, we thus assign an average purchase per user, without losing much information.\\


We observe different types of users. Some people make few big purchases, and others make many small purchases. Our hypothesis is that this difference in behavior predicts household income. The users who make big purchases will be richer than the users who make multiple small ones, simply because the poorer will not have enough cash flow to buy lots of airtime credit all at once. Although there has not yet been a published study that validates this hypothesis, some studies have shown a correlation between amounts of purchases and socio-economic levels \cite{blumenstock2010,blumenstock2012}. Also, carriers have confirmed that, through marketing surveys they have developed reliable predictive models for household income by looking at size and frequency of credit purchases \cite{kirkpatrick_strata}.\\

\begin{figure}
  \centering
  \includegraphics[width=\linewidth]{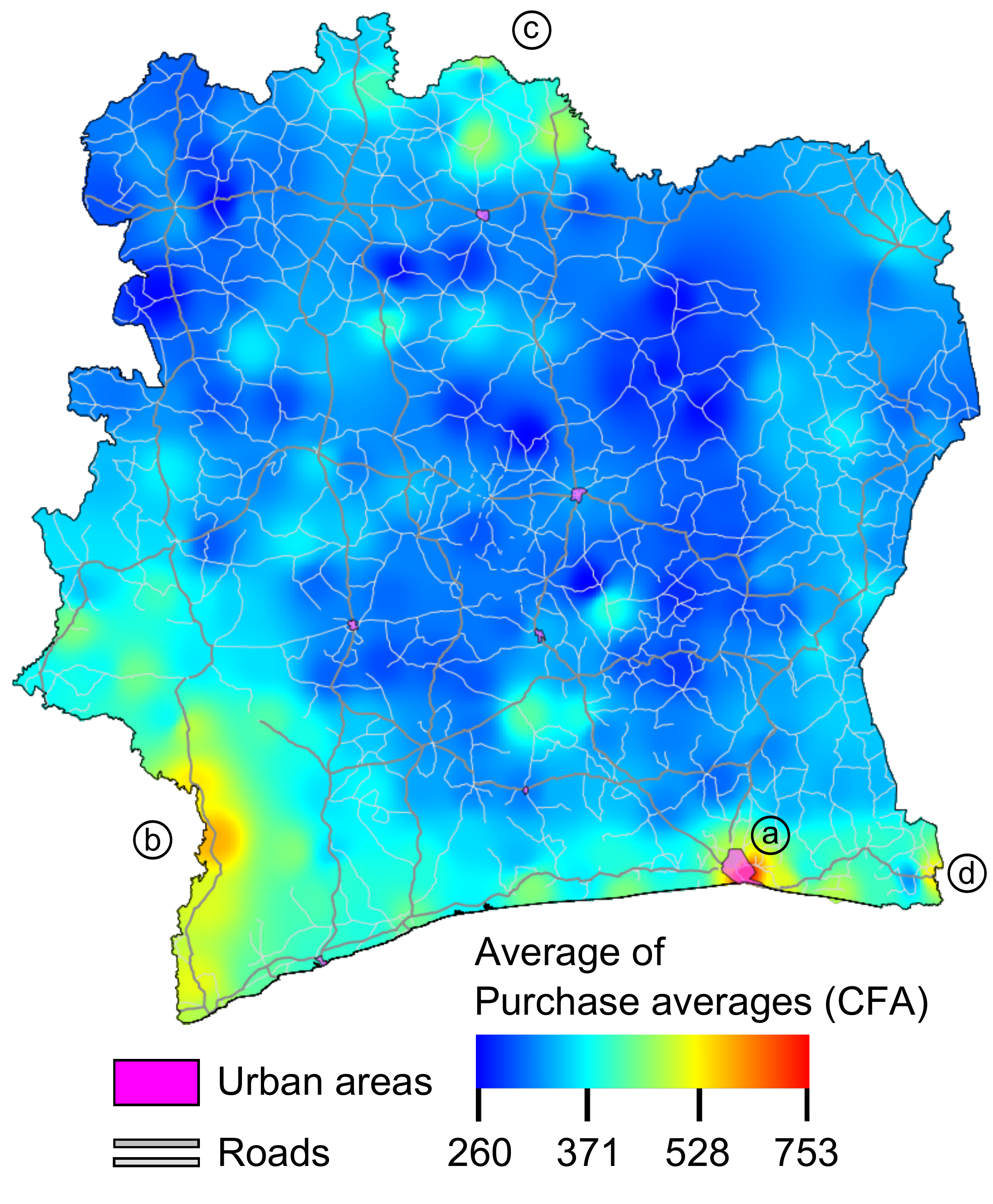}
  \caption{Average of each user's purchase average. (a) Abidjan, (b) Liberian border (c) Roads to Mali and Burkina Faso (d) Road to Ghana}
  \label{fig:topup_mean}
\end{figure}

If we had some up-to-date and reliable data about wealth or household income in the country, we could also build a predictive model of wealth based on size and frequency of purchases. But, as we explain in the Appendix, we do not have such data. Without a training set, we do not have any insight into how to combine or weigh the size and frequency. We can therefore only look at one variable at a time. Between the size and frequency of purchases, we choose to look at the size of purchases, as it reflects the most a user's income or wealth. Someone who is poor will have to buy airtime credit in small amounts while someone who is rich can make larger purchases. We thus map the purchase averages in Figure \ref{fig:topup_mean}.\\

Abidjan has a higher average than most of the country, which was expected with its status of economic capital of Côte d'Ivoire and the biggest sea port in West Africa. On the other hand, the other cities do not have a purchase average that stands out. The Liberian border in the South-West is unexpectedly wealthy. Indeed, the population density is low and there is no apparent reason for the region to be wealthier than another rural area. On the contrary, it is known for its insecurity and land conflicts \cite{fonciers}. However, some of the wealth could come from illegal activities, as it seems to be a location of choice for drug, arms and human trafficking \cite{armsdrug_traf,human_traf}.\\

The borders with the main roads to Mali and Burkina Faso in the North and to Accra (Ghana) in the South-East also stand out. As they are big routes of passage, it is reasonable to expect that these places would stimulate economic activity.\\

Generally the coast in the South is also wealthier, which could be explained by activities linked to tourism.\\

These hypothesis are only speculation : we do not know the causality of these results and it is difficult to interpret them with little information about the context. Nonetheless, this map could be useful as an indicator of potentially interesting further investigations.\\

Furthermore, this map has to be analyzed with care, as low population density areas are more sensitive to outliers. For example, some places seem wealthy, but by looking closer at the unpopulated antenna, it is placed on a big quarry with a dozen of expatriate or executives that completely skew the average.\\

\begin{figure}
  \centering
  \includegraphics[width=\linewidth]{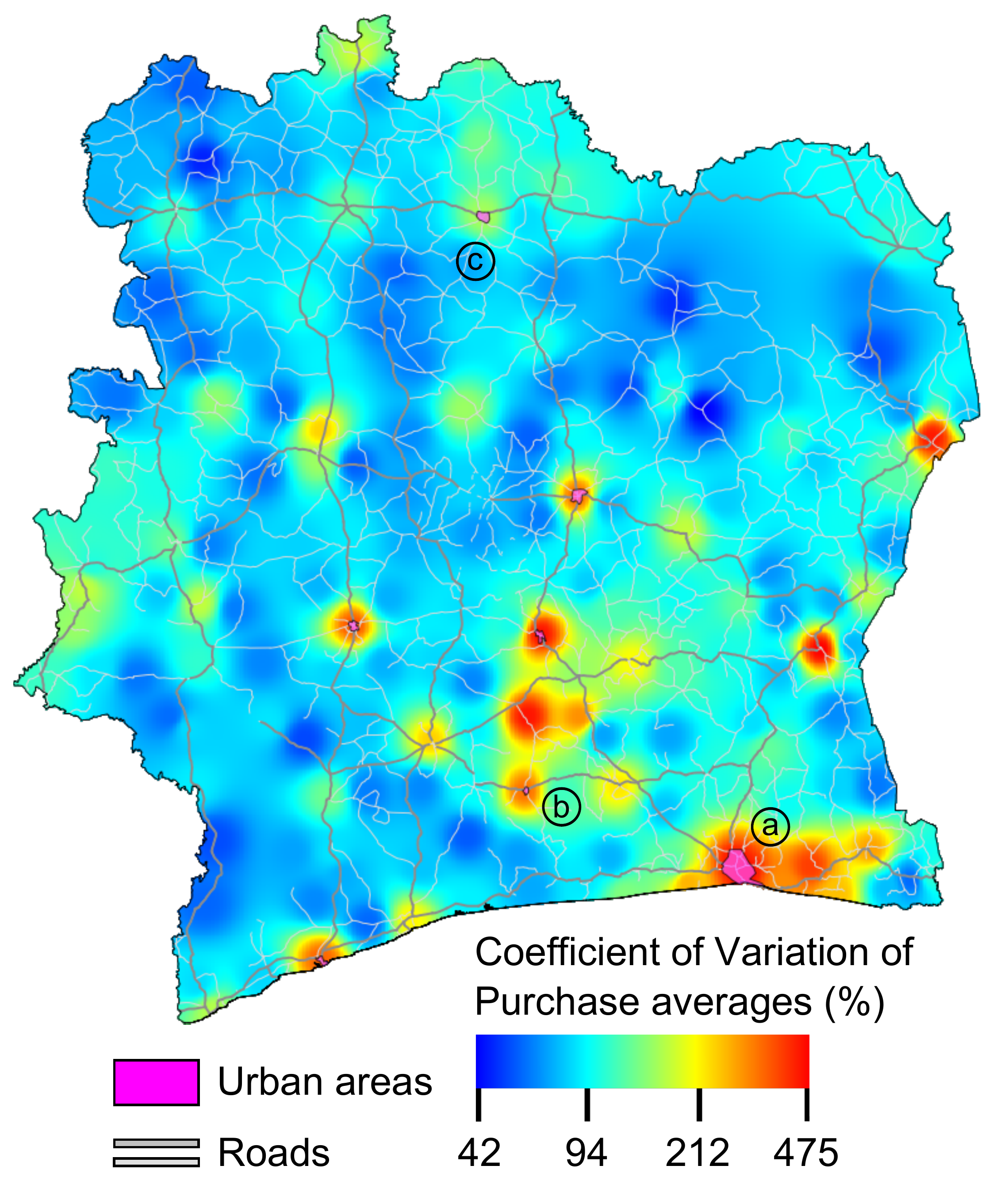}
  \caption{Coefficient of Variation of purchase averages, measures the diversity. (a) Abidjan, (b) Divo (c) Korhogo}
  \label{fig:topup_cv}
\end{figure}

\begin{figure}
  \centering
  \includegraphics[width=\linewidth]{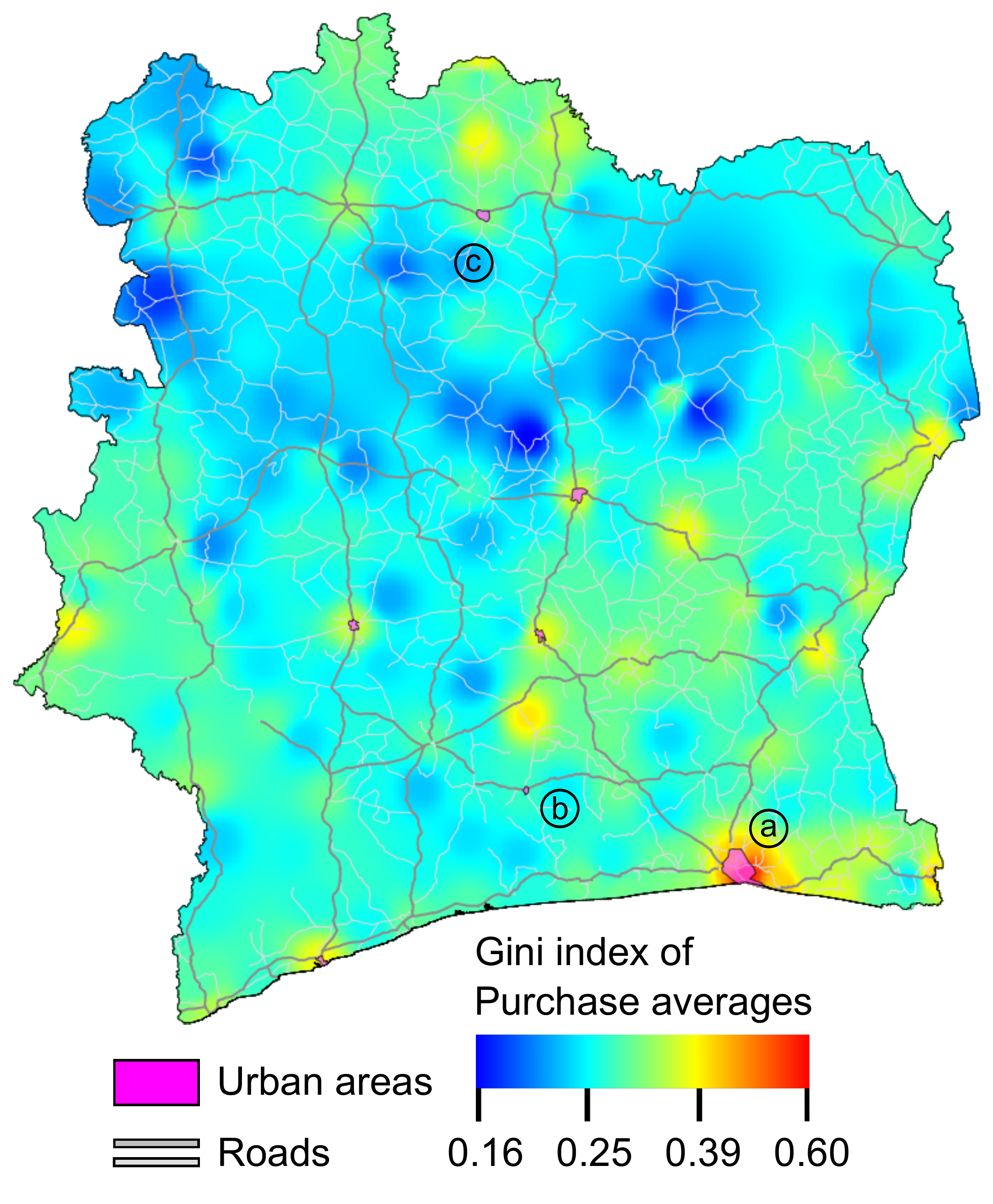}
  \caption{Gini index of purchase averages, measures inequality. (a) Abidjan, (b) Divo (c) Korhogo}
  \label{fig:topup_gini}
\end{figure}

We can map our estimate of wealth diversity and inequality with the Coefficient of Variation and the Gini index of purchase averages, respectively in Figures \ref{fig:topup_cv} and \ref{fig:topup_gini}.\\

Most urban areas clearly stand out in terms of diversity and inequality, confirming that they host poor and rich people. Korhogo, on the other hand, is a city as big as the others \cite{ins} but does not have as much diversity or inequality as them~: people that make small purchases make up a much bigger proportion of its population than in other cities. Divo is almost the same, but its wider mid-class gives it a higher CV. The Liberian border does not have any diversity, indicating that all the people living there do buy purchases in larger amounts.

\section{Social network}

The Call Detail Records tell us who communicates with whom. With this information, we can build the social network of users. Each node in the network corresponds to a client, and two clients are connected if they communicated with each other at least once per month. We do not impose reciprocity, i.e. that there are communications going in both directions. Indeed, we preferred omitting this condition, as imposing it resulted in communities with members living further from each other. We prefer when they live closer to each other, as the center of gravity of communities will then represent the position of its members more accurately, and our mapping will also be more accurate.

Not enforcing reciprocity has some implications, the noise is not as filtered and, therefore, some links do not represent social interaction. The main outliers are hotlines, helpdesks and service numbers which end up in the social network with degrees above the thousands. We identify them with their disproportionate difference between in and out-degree, or when they contacted too many distinct people.\\

Regarding links, they are weighted according to the sum of the number of calls and text messages exchanged between the two users.\\

The resulting network is highly clustered despite its sparseness. Indeed, its average local clustering coefficient, as defined in \cite{local_clustering_new}, is equal to 0.13. On the other hand, the shuffled network (i.e. the network with the same degree distribution, but with links redistributed randomly) presented an average local clustering coefficient of only $2.10^{-5}$. The links within the social network are thus not scattered randomly, and we can detect the communities in the network using the Louvain method \cite{blondel} with a scaled modularity \cite{lambiotte} to detect small group of friends.\\

\begin{figure}
  \centering
  \includegraphics[width=\linewidth]{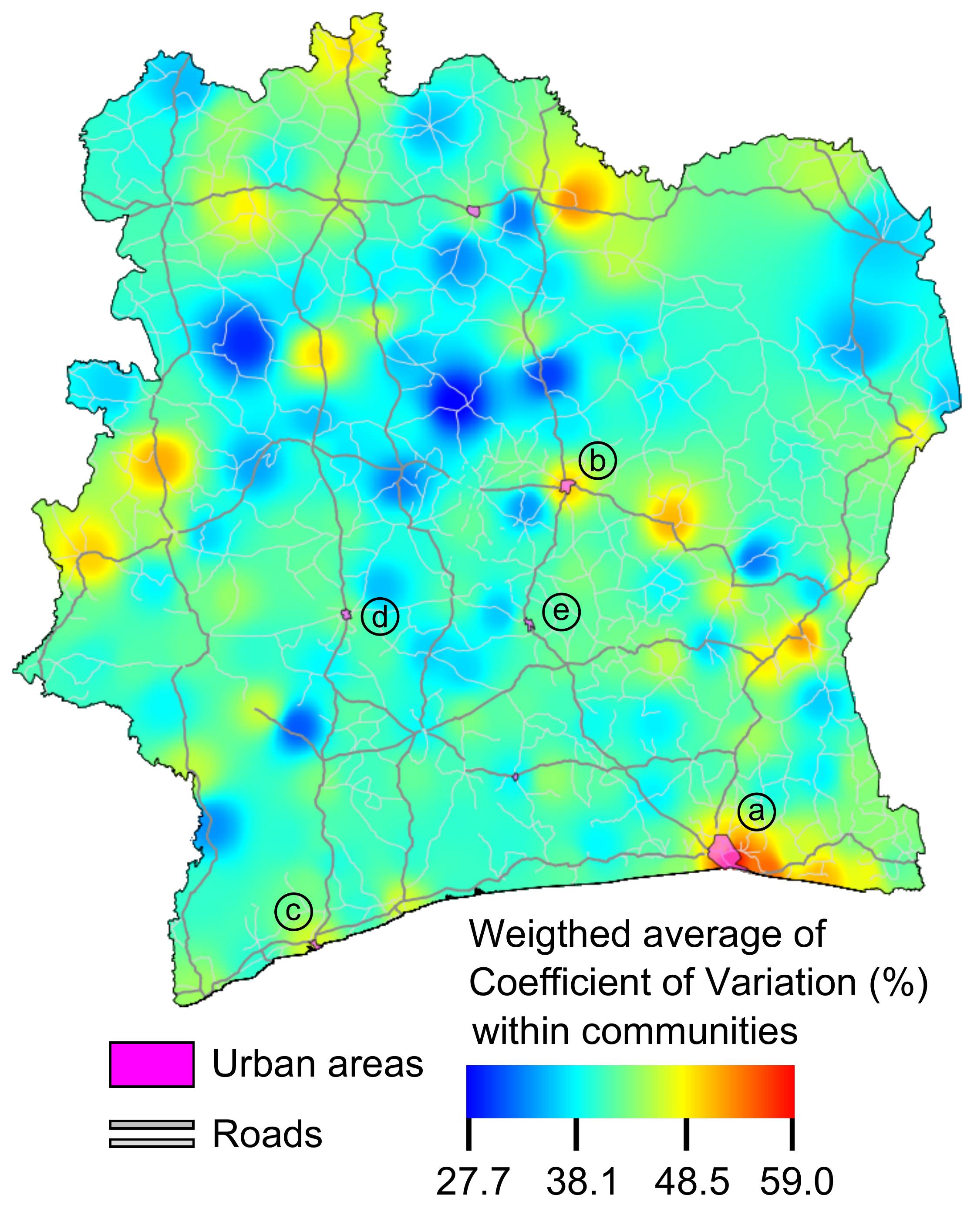}
  \caption{Weighted average of Coefficient of Variation of average purchases within each community. (a) Abidjan, (b) Bouaké (c) San Pédro (d) Daloa (e) Yamoussoukro}
  \label{fig:community_cv}
\end{figure}

Comparing the different members of communities, we find a certain homophilia in terms of purchase averages. Consider the Coefficients of Variation of purchase averages within each community, the weighted average of these values with the size of each community is equal to 48.5\%. If we distribute the purchase averages randomly throughout the social network, we find a weighted average of 68.5\%. This confirms that people tend to be friends with people that have the same purchase average as themselves. In Figure \ref{fig:community_cv}, we map these values to see which places have more diverse communities than others.\\

Some cities tend to have more diverse communities, notably Abidjan, Bouaké and San Pédro. This makes them cities where poor individuals would want to go to, as these are the places where they have the most probability of being part of a community of wealthier people and could, at the same time, climb the socio-economic ladder.\\

Other cities have a high diversity (see Figure \ref{fig:topup_cv}), without having diversity in their communities, giving us a hint of places where socio-economic segregation is stronger than others. Among these cities are Daloa and Yamoussoukro.

\section{Conclusion}

We analyzed airtime credit purchasing behaviors, and noted that most individuals buy credit by amounts of similar sizes. We hypothesize that size and frequency of purchases are correlated with the income of individuals. But, in absence of reliable data to build a model that takes both variables into account, we used the average size of purchases to estimate the relative wealth of an individual. As expected, the economic capital clearly stood out in terms of wealth. Other places that stood out were the coast, with San Pédro, the borders with main roads to Mali, Burkina Faso and Ghana, and all the Liberian border. Other cities do not look wealthier than other rural areas.\\

In terms of diversity, urban areas clearly stood out, with some of them having more inequality than others in the distribution of our proxy for wealth.\\

We detected communities in the social network, and found that people in the same community tend to have the same purchase average. Indeed, the average Coefficient of Variation of purchase average within communities is equal to 48.5\%, compared to 68.5\% when we shuffle the purchase averages.\\

We showed a way to extract indicators of wealth, inequality and segregation out of a mobile phone dataset, and we think that there is a real opportunity to be seized in terms of obtaining low-cost socio-economic statistics from the ubiquitous sensors that are mobile phones. These logs of mobile phone operators can indeed be used to build predictive models, but private or official (but reliable) surveys still need to be conducted to help build more accurate models and assess their quality. Nevertheless, these datasets can be a starting point to understand the socio-economic state of the population in countries with no resources to conduct large surveys, thus helping organizations to make better decisions founded on fact-based statistics.

\begin{figure*}
  \centering
  \includegraphics[width=\linewidth]{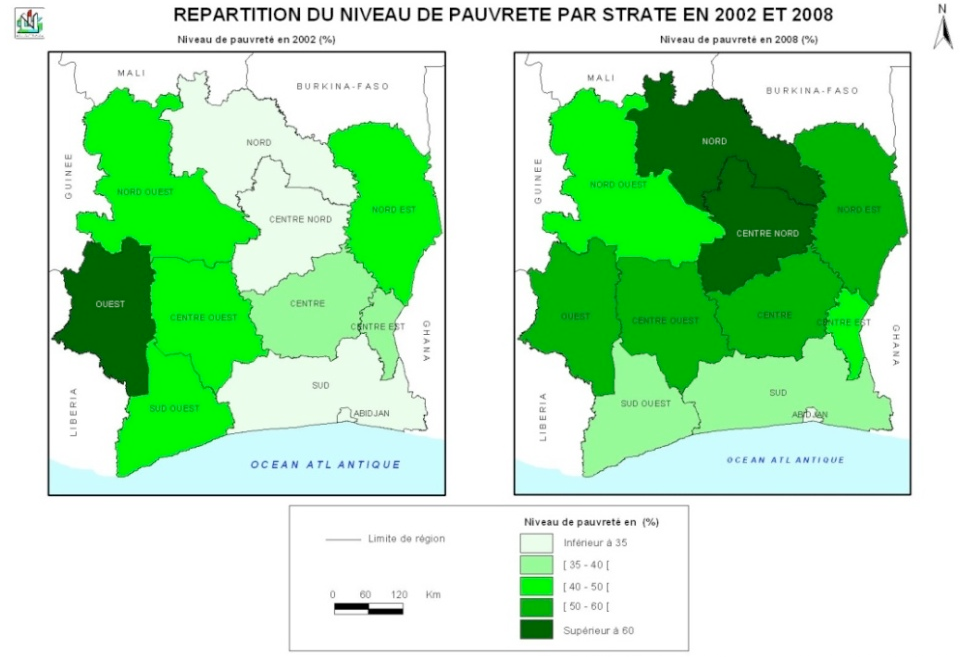}
  \caption{Poverty rates in 2002 and 2008 of Côte d'Ivoire, according to censuses from the National Statistics Office of Côte d'Ivoire. (Realisation : Institut National de la Statistique (INS), Division cartographie)}
  \label{fig:poverty_rates}
\end{figure*}

\appendix[Unreliability of census reports]

We would have liked to compare our data with official reports based on income census like the ones performed by the National Statistics Office of Côte d'Ivoire in 2002 and 2008. The poverty rates of these two reports are given in Figure \ref{fig:poverty_rates}. Unfortunately, we find that either the situation has completely changed in six years' time or the estimated poverty rates are unreliable. If the poverty rates are unreliable, then it is difficult to conclude anything from the comparison. If the situation has changed so much in six years, it is very likely that it has changed a lot in the four years time separating the census and our dataset, even more so with the civil war in between. In this case also, it is difficult to conclude anything from the comparison.

\end{document}